\documentclass[pra,twocolumn,amsmath,amssymb,groupedaddress,superscriptaddress]{revtex4}
\usepackage{graphicx}

\begin{document}

\def\RR{\mathbb{R}}
\def\S{{\cal S}}
\def\T{{\cal T}}
\def\dd{\delta}
\def\one{{\bf 1}}
\def\kk{{\bf k}}
\def\KK{{\bf K}}
\def\qq{{\bf q}}
\def\rr{{\bf r}}
\def\nn{{\bf n}}
\def\uu{{\bf u}}
\def\mm{{\bf m}}
\def\LL{{\rm L}}
\def\AA{{\rm A}}
\def\BB{{\rm B}}
\def\aat{{\rm at}}
\def\pph{{\rm ph}}
\def\vvac{{\rm vac}}
\def\iint{{\rm int}}
\def\00{{\bf 0}}
\def\RRe{{\rm Re}}
\def\IIm{{\rm Im}}
\def\ccoh{{\rm coh}}
\def\iinc{{\rm inc}}
\def\TTr{{\rm Tr}}

\title{Collective generation of quantum states of light by entangled atoms}

\author{D. \surname{Porras}}
\email{Diego.Porras@mpq.mpg.de}
\author{J.~I. \surname{Cirac}}
\email{Ignacio.Cirac@mpq.mpg.de}
\affiliation{
Max-Planck-Institut f\"ur Quantenoptik, Hans-Kopfermann-Str. 1, Garching, D-85748, Germany.}

\begin{abstract}
We present a theoretical framework to describe the collective emission of light
by entangled atomic states.
Our theory applies to the low excitation regime, where most of the
atoms are initially in the ground state, and relies on a
bosonic description of the atomic excitations.
In this way, the problem of light emission by an ensemble of atoms can be solved
exactly, including dipole-dipole interactions and multiple light scattering.
Explicit expressions for the emitted photonic states
are obtained in several situations,
such as those of atoms in regular lattices and atomic vapors.
We determine the directionality of the photonic beam, the purity of
the photonic state, and the renormalization of the emission rates.
We also show how to observe collective phenomena with ultracold atoms in optical lattices, and
how to use these ideas to generate photonic states that are useful in
the context of quantum information.
\end{abstract}

\date{\today}

\maketitle

\section{Introduction}
In the last years, the field of atomic, molecular and optical physics
has witnessed an impressive advance in the development of setups
to trap atoms under different conditions, like for example, ions in electromagnetic traps
and ultracold atoms in optical lattices.
Furthermore, the quantum state of these systems may be engineered by
performing quantum operations such as as quantum gates between ions
\cite{ions.gates}, or the excitation of neutral atoms under the dipole blockade
\cite{LukinRydbergatoms,exp.Rydberg}.
In this way one can create deterministically collective entangled states, like
the completely symmetric states with a single excited atom
($| W \rangle$-states) \cite{Häffner}.
In addition, those are systems where atoms can
be coupled to light in a very controlled way. Since some of the atomic
entangled states which may be created in those setups play an
important role in the description of the interaction of atomic
ensembles with light,
an important question arises, namely, can we use our control on the states of trapped atoms
to generate useful quantum states of light? If so, which are the
properties of such states in terms of photon directionallity, purity,
or photon entanglement?

A preliminary study of these ideas was recently presented in
\cite{preprint}. The understanding of this problem
requires a theoretical framework to
describe the emission of light by collective atomic states under a variety
of trapping conditions.
Related problems have been indeed a subject of 
investigation since the seminal work by Dicke
\cite{Dicke}. Nevertheless, the experimental systems that we have in mind
share a few peculiarities which demand a new approach for their
description.
First, ultracold atoms are trapped in ordered arrangements such as Coulomb crystals or
optical lattices, where the distance between atoms, $d_0$, is comparable to the
wavelength of the light, $\lambda$.
This situation is very different from Dicke
superradiance, where  $\lambda$ is larger than the size of the whole
system. It also differs from the case of crystals, where $\lambda \gg
d_0$ as considered, for example, in \cite{Scully06}.
Also, we study a situation in which atoms are initialized in a
collective state, which is not necesarilly created by
the absorption of a photon. Thus, rather than the more traditional
description in terms of light scattering, we need
a theoretical model of the mapping
between atomic collective excitations and photons.

In its more general form the above described situation poses a very
complicated many-body problem.
A crucial simplification is achieved by considering the low
excitation sector of the atomic Hilbert space, which in the Holstein-Primakoff (HP)
approximation can be described in terms of bosonic spin-waves
\cite{HP}. This approach has been used in previous works, for example,
to study Dicke superradiance \cite{Ressayre}, slow propagation of
light \cite{Lukin00}, and atom-light interfaces with atomic vapors \cite{Hammerer}.
Also, the
emission of light by ensembles of harmonic oscillators was studied in
\cite{Zakowicz}, although in a different regime of
trapping conditions than those considered here.

In this paper we make use of the HP approximation to describe the collective
emission of light as a mapping between spin-wave excitations and
photons. For atoms placed at fixed positions this mapping is a
Gaussian completely positive map \cite{Giedke}, and we present a
method to get explicit expressions of the photonic modes into which light is emitted.
Furthermore, we extend this formalism to study the case of
atomic vapors. We obtain the following results:
(i) For atoms trapped in a regular lattice, there is a regime in
which photons may be emitted in a collimated beam, which requires that
$\lambda > 2 d_0$. In this regime, there is a renormalization of the emission rates,
leading to a classification of the
low-excitation atomic Hilbert space in terms of
superradiant and subradiant spin-waves. Superradiant states decay
with a rate that is enhanced by a factor $\chi$, which depends on the
dimensionality of the lattice.
In 1D and 3D we determine the values $\chi_{1D} \propto \lambda/d_0$ and
$\chi_{3D} \propto (\lambda/d_0)^2 (L / d_0)$, respectively, with $L$
the length of the lattice. This effect is related but not equivalent
to Dicke superradiance.
(ii) In the case of atomic vapors the
collective characeter in the emission of light is determined by
$\chi_{\rm en} \propto N (\lambda/L)^2$, where $N$ is the total number of
atoms.  The directional regime requires $\chi_{\rm en} \gg 1$, and the
emission rate is enhanced by $\chi_{\rm en}$.
(iii) Some of these effects, like the 
renormalization of the emission rates and the directionality, 
could be observed
in a relatively simple experiment with atoms in optical lattices.
(iv) By making use of coherent effects,
photonic entangled states could be generated by trapped ions or
neutral atoms. Photons that are generated in this way could be collimated and in a
pure state, and thus be useful in the context of quantum information.
\section{Theoretical framework}
Our first task is to describe the collective emission of light by an
ensemble of atoms. First, we focus on the situation in which atoms are
placed at fixed positions. At the end of the section we discuss the
effect of atomic motion.

\subsection{The atom-photon map in the Holstein-Primakoff approximation}
\begin{figure}[h]
\resizebox{!}{!}{\includegraphics{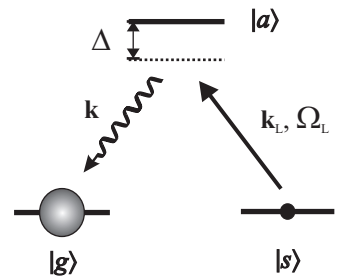}}
\caption{Lambda scheme.}
\label{fig.lambda}
\end{figure}

We consider an ensemble of $N$ atoms trapped by harmonic
potentials with trapping frequency $\nu$, and internal levels forming a
$\Lambda$-scheme, see Fig. \ref{fig.lambda}.
Two ground states $| g \rangle$, $| s \rangle$, are
coupled by means of a laser with Rabi frequency $\Omega_\LL$ and
wavevector $\kk_\LL$, through and auxiliary level $| a \rangle$.
We define atomic operators
$\sigma_j = | g \rangle_j \langle s |$,
$\sigma^{\rm{ss}}_j = | s \rangle_j \langle  s |$, which fulfill the
commutation relations,
\begin{equation}
\langle [\sigma_j , \sigma_l^+]  \rangle =
\delta_{j,l} \left( 1 - 2 \langle \sigma^{\rm ss}_j \rangle  \right).
\label{HP.approximation}
\end{equation}
Under the condition that the excitation probability of each atom is low,
$\langle \sigma_j^{\rm ss}  \rangle \ll 1$, we replace atomic
operators by HP bosons, $\sigma_j \rightarrow b_j$.

The HP approximation allows us to recast the atom-light Hamiltonian as
a bosonic quadratic Hamiltonian. After the adiabatic elimination of the upper
level $| a \rangle$, our system is described by
\begin{equation}
H = H_0 + H_{\rm lm} + H_{\rm mot} .
\label{total.Hamiltonian}
\end{equation}
$H_0 = \sum_\kk \omega_k a^+_\kk a_\kk$, with $\omega_k = c k$.
We consider, for the shake of clarity, a scalar model for the electromagnetic
field, since our conclusions do not change when including the dipole
pattern, as we show later. However, the photon polarization can be
included straightforwardly in our formalism.
$H_{\rm lm}$ is the atom-light interaction Hamiltonian in the rotating
wave approximation \cite{note.rwa} (we set $\hbar
= 1$),
\begin{eqnarray}
H_{\rm lm}
&=& \sum_{j, \kk} g_k b_j^+  a_{\bf k} e^{i (\kk - \kk_\LL) {\bf r}_j +
  i \omega_\LL t}  +  \textmd{h.c.} \ , \nonumber \\
g_k &=& \frac{\Omega_\LL}{2 \Delta} \frac{\hbar \omega_k}{2 \epsilon_0
V} d_{\rm ga} .
\label{H.lm}
\end{eqnarray}
$V$ is the quantization volume of the
electromanetic field, $\epsilon_0$ the vacuum permitivity, and
$d_{ga}$ the dipole matrix element of the $|g\rangle$-$|a\rangle$ transition.
$H_{\rm mot}$ is the Hamiltonian describing the motion of the
atoms during the emission process, whose energy scale is given
typically by the trapping frequency $\nu$. Eq. (\ref{H.lm}) is
obtained under the assumption $\Delta \gg \Omega_\LL, \nu \bar{n}$,
with $\bar{n}$ the mean vibrational occupation number.
In the resonant case, $\Delta = 0$, 
the validity of the adiabatic elimination of 
$|a\rangle$ requires that the decay rate from $|a\rangle$ to
$|g\rangle$, $\Gamma_{\rm ag}$, satisfies 
$\Gamma_{\rm ag} \gg \Omega_\LL, \nu \bar{n}$, 
and a similar coupling as (\ref{H.lm}) is obtained.
The spontaneous decay of $|a\rangle$ back to $|s\rangle$ is neglected
in the forecoming analysis. This is justified either when collective effects
enhance the $|e\rangle$-$|g\rangle$ channel,
or by choosing $|e\rangle$-$|g\rangle$ to be a cycling transition, see
Appendix \ref{appendix.cycling}.
We deal first with the simplest case, in which atoms
are placed at fixed positions and the motion of the atoms can be neglected.
This is a good approximation in the Lamb-Dicke regime, $(k_\LL x_0)^2
(2 \bar{n} + 1) \ll 1$, where $x_0$ is the ground state size in the
harmonic trap. 
The effect of the motion is considered
in the last subsection.
This work relies on the observation
that $H_0 + H_{\rm lm}$ generates a  beam-splitter
transformation between atomic states and photons. The evolution of the
atom-photon system can thus be understood in terms of a Gaussian
completely positive map \cite{Giedke}. In the low
excitation regime, any atomic state can be expressed in terms of
bosonic excitations,
\begin{equation}
| \Psi \rangle_\aat = \sum_{n_1, \dots, n_N}
\Psi_{n_1, \dots, n_N}
\frac{(b^+_1)^{n_1}}{\sqrt{n_1 !}} \dots
\frac{(b^+_N)^{n_N}}{\sqrt{n_N !}}
| 0 \rangle_\aat .
\label{Psi}
\end{equation}
We consider the following initial state,
\begin{equation}
| \psi(0) \rangle = | \Psi \rangle_\aat | 0 \rangle_\pph ,
\label{initial.state}
\end{equation}
where $| 0 \rangle_\pph$ is the photon vacuum.
Our goal is to find the photon state $| \Phi \rangle_\pph$ at a time
$t$ longer than the atomic decay time,
\begin{equation}
| \psi(t) \rangle = {\cal U}(t) | \psi(0) \rangle =
| 0 \rangle_\aat | \Phi \rangle_\pph ,
\label{final.state}
\end{equation}
where ${\cal U}(t) = e^{-i H t}$, and $H$ is the total atom-system Hamiltonian.
Together with the beam-splitter form of
(\ref{H.lm}) this implies that the problem is reduced to finding the
exact form of the transformation,
\begin{equation}
{\cal U}(t) b_j^+ {\cal U}(t)^\dagger = \sum_\kk g_{j \kk}(t) a^+_\kk +
\sum_l h_{j l}(t) b^\dagger_l .
\label{c.transformation}
\end{equation}
Taking into account that ${\cal U}(t)^\dagger | \vvac\rangle = 0$,
with $| \vvac \rangle = |0\rangle_\aat |0\rangle_{\rm ph}$, then
$g_{j \kk}(t)$ can be found by using the relation,
\begin{equation}
g_{j \kk}(t) = \langle \vvac  | a_\kk {\cal U}(t) b_j^+  | \vvac \rangle =
\langle \vvac | a_\kk (t) b_j^+ | \vvac \rangle ,
\label{key.observation}
\end{equation}
where $a_\kk(t)$ is
the photon operator in the Heisenberg picture. Also,
\begin{equation}
h_{j l}(t) = \langle \vvac  | b_l {\cal U}(t) b_j^+  | \vvac \rangle =
\langle \vvac | b_l (t) b_j^+ | \vvac \rangle ,
\label{key.observation.2}
\end{equation}
which shows that $h_{j l}(t) = 0$ for $t$ longer than the atomic decay
time.
$g_{j \kk}(t)$ is thus the only interesting term, since it determines the
photon mode into which light is radiated. Its determination can be
readly done by means of the Heisenberg equation of motion for the
photonic operators, which yields
\begin{eqnarray}
&& g_{j \kk}(t) =  \label {g.jk} \\ \nonumber
&& - i g_k  e^{-i\omega_k t} \sum_l \int_0^t  d \tau e^{- i (\kk - \kk_\LL) \rr_l + i
(\omega_k - \omega_\LL) \tau}
\\ \nonumber
&& \hspace{5cm}
\langle \vvac | b_l(\tau) b^+_j | \vvac \rangle . \nonumber
\end{eqnarray}
This equation is our starting point for the exact determination of the
photonic modes.
\subsection{Photonic modes}
To find the mapping between atomic and photonic modes,
we need to solve the master equation, to calculate the time
evolution of the atomic correlator in Eq. (\ref{g.jk}).
In general, this is a difficult many-body problem,
but in the bosonic limit it can be described exactly. In terms of HP
bosons, the master equation which describes the atomic dynamics reads 
\cite{Lehmberg},
\begin{eqnarray}
\frac{d \rho}{d t} &=&
\sum_{i,j}
\left( J_{ij} b_j \rho b^+_i - J_{ij} b^+_i b_j \rho + \textmd{h.c.} \right) ,
\label{master.eq}
\end{eqnarray}
where the $J_{ij}$ include multiple light scattering and
dipole-dipole interactions,
\begin{equation}
J_{ij} =
\sum_\kk g_k^2 \int_0^\infty
e^{i (\omega_k - \omega_\LL) \tau + i ({\bf k} - {\bf k}_\LL)(\rr_i -
  \rr_j)} d \tau .
\label{master.eq.coeff.0}
\end{equation}
This expression is evaluated by using the identity $\int d \tau e^{i
  \omega \tau} = \pi \delta(\omega) + i {\cal P}(1/\omega)$, which
yields the result,
\begin{eqnarray}
J_{i i} &=& \frac{1}{2} \bar{\Gamma} , \\
J_{i j} &=& \frac{1}{2} \bar{\Gamma} e^{-i \kk_\LL (\rr_i - \rr_j)}
\nonumber \\
& & \hspace{0.5cm}
\left( \frac{\sin(k_\LL |\rr_i - \rr_j|)}{k_\LL |\rr_i -\rr_j|}
- i \frac{\cos(k_\LL |\rr_i - \rr_j|)}{k_\LL |\rr_i - \rr_j|}
\right) (i \neq j) , \nonumber
\end{eqnarray}
where
\begin{equation}
\bar{\Gamma} = \frac{1}{3\pi} 
\left( \frac{\Omega_\LL}{2 \Delta} \right)^2
\frac{\omega_\LL^3}{\epsilon_0 c^3} d_{\rm ga}^2 ,
\label{master.eq.coeff}
\end{equation}
is the single atom decay rate.
The inclussion of the photon polarization would change the spatial
dependence of the couplings $J_{i j}$.
However, as we show later, the
collective phenomena would be the same. Note that we
are not including the single atom Lamb shift, which may be simply
absorved into the laser frequency $\omega_\LL$.
Since Eq. (\ref{master.eq}) is quadratic in bosonic operators, it is readly
solved by defining eigenmodes which diagonalize the atomic
quantum dynamics in the low-excitation limit,
\begin{eqnarray}
&& b_j  = \sum_n {\cal M}_{j n} b_n ,
\nonumber \\
&& \left({\cal M}^{-1} J {\cal M} \right)_{n m} = J_n \delta_{n,m}.
\label{master.eigenstates}
\end{eqnarray}
The matrix $J_{ij}$ is not hermitean, and thus canonical commutation
relations are not conserved, $[b_n,b^+_m] \neq \delta_{n,m}$. However,
the evolution of averages takes a simple form given by
\begin{equation}
\langle b_l(\tau) \rangle = \sum_n {\cal M}_{l n} e^{- J_n
  \tau} \langle b_n(0)  \rangle .
\label{average.evolution}
\end{equation}
Thus, the spin-wave dynamics is governed by the eigenvalues  $J_n$. 
The latter contain the collective decay rates, $\Gamma_n$, and the collective energy shifts, 
\begin{equation}
\Gamma_n = 2 \ {\rm Re}\left( J_n \right), \hspace{0.2 cm}
\Delta_n = 2 \ {\rm Im} \left(J_n \right).
\end{equation}
Conservation of the trace under the transformation
(\ref{master.eigenstates}) leads to the following sum rules,
\begin{equation}
\sum_n \Gamma_n = N \bar{\Gamma}, \hspace{0.2cm} \sum_n \Delta_n = 0.
\end{equation}
Note that whenever collective effects
induce a renormalization of $\Gamma_n$, the sum rule implies the
existence of super- and subradiant states.

By application of (\ref{master.eigenstates}, \ref{average.evolution}),  and
the quantum regression theorem, we determine the two-time atomic
average in (\ref{g.jk}). 
In the limit that $t \gg 1/\Gamma_n$ we get
\begin{eqnarray}
g_{j \kk}(t) =
i e^{-i \omega_k t} g_\kk \sum_{l n}
\frac{
\left( {\cal M}^{-1} \right)_{n j}
e^{-i (\kk - \kk_\LL) \rr_l} {\cal M}_{l n}}
{i (\omega_k - \omega_\LL) - J_n} ,
\label{mapping}
\end{eqnarray}
which yields the explicit form of the atom-photon
mapping. By means of this relation it is possible to determine the
many-photon state emitted by any initial atomic state like
(\ref{Psi}). 

To get insight of the characteristics of the photonic
states emitted by the collective atomic states, we focus from now on, 
on the mapping to a single photon.
To clarify the notation, let us define $| \Psi_n \rangle_\aat$, the
$n$-spin-wave state with a single excitation,
\begin{eqnarray}
| \Psi_n \rangle_\aat
&=& \frac{1}{{\cal N}_n}
\sum_j {\cal M}_{j n} b^+_j | 0  \rangle_\aat ,
\nonumber \\
{\cal N}_n &=& \sqrt{\sum_j {\cal M}^*_{j n} {\cal M}_{j n}} .
\label{collective.state}
\end{eqnarray}
${\cal N}$ has to be included due to the non-hermiticity of $J_{ij}$.
We determine $|\Phi_{n}\rangle_{\pph}$, the single photon state into which 
$| \Psi_n \rangle_{\aat}$ is mapped,
\begin{eqnarray}
| \Phi_n(t) \rangle_\pph &=& \sum_\kk \phi_{n,\kk}(t) a^+_\kk | 0 \rangle_\pph  ,
\nonumber \\
\phi_{n,\kk}(t) &=& i \frac{g_k e^{-i \omega_k t}}{{\cal N}_n} \sum_l
\frac{e^{-i (\kk - \kk_\LL) \rr_l} {\cal M}_{l n}}{i (\omega_k -
  \omega_\LL) - J_n} .
\label{photon.state.n}
\end{eqnarray}
Although these results allow one to solve exactly the problem of
collective emission of light including the effects of reabsorption,
aditonal insight can be gained by considering the case of a system
with periodic boundary conditions. This will be a good approximation
for a finite system, provided that the number of atoms in the volume
is much larger than in the surface, that is, $L/d_0 \gg 1$. In this
case we get the matrices
\begin{equation}
{\cal M}_{j n} = \frac{1}{\sqrt{N}} e^{i {\bf K}_n \rr_j} .
\label{plane.wave}
\end{equation}
The spin-wave state $n$ is then defined like
\begin{equation}
| \Psi_n \rangle_\aat = \frac{1}{\sqrt{N}} \sum_j e^{i {\bf K}_n
  \rr_j} b^+_j | 0 \rangle_\aat .
\label{atomic.plane.wave}
\end{equation}
The vector ${\bf K}_n$ is the momentum of the collective atomic
state. The resulting photonic mode is defined by the following expression,
\begin{equation}
\phi_{n,\kk}(t)  =
i g_k e^{-i \omega_k t}
\frac{\frac{1}{\sqrt{N}}\sum_l e^{- i (\kk - \kk_\LL - {\bf K}_n)
    \rr_l}}{i (\omega_k - \omega_\LL) - J_n} ,
\label{mapping.pbc}
\end{equation}
such that collective effects and dipole-dipole interactions enter
through the dependence of the collective emission rate $J_n$ on the
mode number, $n$. Note that by using periodic boundary conditions,
the matrix ${\cal M}$ define an unitary transformation, and thus the sets
${| \Psi_n \rangle}_{n = 1,\dots,N}$, and ${| \Phi_n
  \rangle}_{n=1,\dots,N}$, form an orthogonal basis of spin-waves, and
photonic modes, respectively.

\subsection{Angular photon number distribution}
We determine now the properties of the emitted photonic modes.
In particular, let us define
\begin{equation}
I(\Omega) =  \frac{V}{(2\pi)^3} \int_0^\infty \langle a^+_\kk a_\kk \rangle k^2 dk ,
\label{def.angular}
\end{equation}
the average photon number per solid angle. Consider an initial
atomic state with a single excitation,
\begin{equation}
| \psi \rangle_\aat = \sum_j \psi_j b^+_j | 0 \rangle_\aat .
\label{define.collective.atomic}
\end{equation}
The emitted photon distribution is
\begin{eqnarray}
I(\Omega) &=& \sum_{j,j'} \psi_j^* I_{j j'} \left( \Omega \right)
\psi_{j'}, \nonumber \\
I_{j j'} \left( \Omega \right) &=& \frac{V}{(2 \pi)^3}
\int_0^\infty g^*_{j, \kk} g_{j', \kk} k^2 dk .
\label{define.matrix}
\end{eqnarray}
Upon substitution of (\ref{mapping}) in the expression for
$I_{j j'}(\Omega)$, and under the condition that $\Gamma_n / c \ll
1/L$ (see Appendix \ref{appendix.propagation}), we get,
\begin{eqnarray}
I_{j j'} \left( \Omega \right)
&=&
\frac{\bar{\Gamma}}{4 \pi}
\sum_{n, n'}
B(\Omega)_{j n} \frac{1}{\bar{J}^*_n + \bar{J}_{n'}} B(\Omega)_{j' n'},
\nonumber \\
B(\Omega)_{j n} &=& \left( {\cal M}^{-1} \right)^*_{n j}
\sum_l {\cal M}^*_{l n} e^{i (k_\LL \nn_\Omega - \kk_\LL) \rr_l} ,
\label{photon.distribution}
\end{eqnarray}
where $\uu_\Omega$ is a unit vector pointing in the direction of the
solid angle $\Omega$. In spherical coordinates, 
$\Omega$ is determined by $(\theta, \phi)$, such that
$\uu_\Omega = (\sin \theta \cos \phi, \sin \theta \sin \phi, \cos \theta)$.

The general recipe for calculating the photon distribution probability
involves the following steps: (i) Calculate the coefficients of
the master equation and find the eigenvalues and eigenvectors of
$J_{ij}$. (ii) Use Eq. (\ref{photon.distribution}) to calculate
$I_{ij}$. 
(iii) Use the latter to calculate the emission spectrum with the
wavefucntion of any given initial atomic state expressed in terms of
bosonic spin-waves. In the planewave approximation
(\ref{plane.wave}) we can get close expressions for the photon
distribution. We focus again in the single photon case, and 
define $I_n(\Omega)$ as the photon number distribution corresponding
to the photonic mode emitted by $|\Psi_n \rangle_\aat$. By using Eq. 
(\ref{photon.distribution}) we get
\begin{equation}
I_n \left( \Omega \right) =
\frac{1}{4 \pi} \frac{\bar{\Gamma}}{\Gamma_n}
\sum_{j,j'} e^{i(\kk_\Omega - \kk_\LL- \KK_n)(\rr_j - \rr_{j'})}.
\label{photon.distribution.collective}
\end{equation}
This result has a clear interpretation in terms of interference of
light emitted by the atomic system \cite{Jackson}. 
Eq. (\ref{photon.distribution.collective}) not only
allows one to calculate the angular emission probabilty, but also, due
to the normalization condition $\int I_n(\Omega) d \Omega = 1$,
\begin{equation}
\frac{\Gamma_n}{\bar{\Gamma}}
= \int d \Omega  \frac{1}{4 \pi}
\sum_{j,j'} e^{i(\kk_\Omega - \kk_\LL- \KK_n)(\rr_j - \rr_{j'})}.
\label{collective.rates}
\end{equation}
This expression provides us with a simple way to determine the collective rates
under the planewave approximation.

\subsection{Effects of the atomic motion}
\label{effects.motion}
Finally, we discuss the effect of the atomic motion on the light
emission. In the most general case, the inclusion of the motional degrees of
freedom poses a very complicated problem which goes beyond the scope
of this work.
Two time scales determine
this problem. First, $\tau_{\rm mot}$,
the time scale of the motion of atoms in the trap.
In the case of trapped particles,
$\tau_{\rm mot} = 1/\nu$, with $\nu$ the trapping frequency. 
We can extend this discussion to the case of atomic vapors, and
consider that in this case, $\tau_{\rm mot} = L/v$, with $L$ the
length of the sample, and $v$ the atom velocity.
$\tau_{\rm mot}$ is to be compared with the radiative decay time, $1/\bar{\Gamma}$,
or more specifically, the set of collective decay times $1/\Gamma_n$. Based on the
comparison between these time scales we define two limits in which
the application of our theoretical framework is particularly straightforward.

{\it (i) Slow motion limit},
$\tau_{\rm mot} \gg 1/\Gamma_n$.
Since the emission
process is much faster than the motion of the particles, we can assume that
atomic positions are frozen. The system is in the initial state
\begin{equation}
| \psi(0) \rangle = | \Psi \rangle_\aat | 0 \rangle_\pph | \Psi_{\bf m}
\rangle_{\rm mot} ,
\end{equation}
where $\Psi_{\rm m} (\rr_1, \dots, \rr_N)$ is the initial wavefunction
in terms of the atomic positions,
which in this limit does not evolve
during the emission time. The atom-photon mapping can be still applied
to this system by solving it for each value of the atomic positions,
\begin{equation}
| \Psi \rangle_\aat | 0 \rangle_\pph
\to
| 0 \rangle_\aat | \Phi^{\rr_1,\dots,\rr_N} \rangle_\pph ,
\end{equation}
where $| \Phi^{\rr_1,\dots,\rr_N} \rangle_\pph$ is the photonic state
obtained under the assumptions that ions are located at positions
$\rr_1,\dots \rr_N$. Any photonic observable, $O$, is then obtained upon
averaging with respect to the wavefunction $\Psi_{\rm m}$, for
example,
\begin{equation}
\langle O \rangle = \sum_{\rr_1,\dots,\rr_N}
\langle \Phi^{\rr_1,\dots,\rr_N} | O  | \Phi^{\rr_1,\dots,\rr_N} \rangle |\Psi(\rr_1,\dots,\rr_N)|^2.
\end{equation}
This method can be readly extended to the case of a mixed motional
state.

{\it (ii) Fast motion limit}, $\tau_{\rm mot} \ll 1/\Gamma_n$.
In this case, we are assuming that trapped atoms move along the sample
in a time that is smaller than the emission time. This case is
particularly relevant, since it describes hot atomic
vapors. We notice first that condition $\tau_{\rm
  mot} \ll 1/\Gamma_n$, implies that the atomic positions are not
correlated with the atomic operators in Eq. (\ref{g.jk}). Thus, we can
describe the atomic radiative decay indepently of the evolution of
atomic positions $\rr_l$ in (\ref{g.jk}). This can be done by using a
master equation with averaged coefficients, 
\begin{equation}
J_{ij}
=
\pi \sum_\kk g_k^2 \
\delta(\omega_k - \omega_\LL)
\langle e^{i ({\bf k} - {\bf k}_\LL)(\rr_i - \rr_j)} \rangle_{\rm mot},
\label{master.eq.coeff.average}
\end{equation}
In the case of an harmonic trap, $\rr_i$ are position operators, and
$\langle \dots \rangle_{\rm mot}$ is the average with the atomic motional
state. This situation can be extended to describe hot atomic vapors,
by replacing the atomic positions, $\rr_i$,
by a set of random variables with a given probability
distribution $\rho_{\rm m}(\rr_1,\dots,\rr_N)$, and performing the
corresponding average to get $J_{ij}$.

Once the atomic dynamics is solved, our results on the atom-photon
mapping can be used to describe the emitted photons.
A general atomic
state is mapped now into a mixed photonic state.
For example, consider the atomic state $|\Psi_m \rangle_\aat$,
defined by Eq. (\ref{collective.state}), which is obtained
by the diagonalization of the
master equation with (\ref{master.eq.coeff.average}). The atom-photon
mapping yields the following photon density matrix after the emission process,
\begin{equation}
\rho^{\pph}(t)  = \sum_{\kk,\kk'} \langle \phi_{n,\kk}(t) \phi_{n,\kk'}^*(t)
\rangle_{\rm mot} a^+_\kk | 0 \rangle_\pph \langle 0 | a_\kk ,
\label{d.matrix}
\end{equation}
and this expression is easily generalized to the multiphoton case.
Note that the solution of the fast motion limit seems similar to the case of slow
motion. However, the crucial difference is that in the fast case, we
are allowed to solve the radiative emission problem, and to perform
subsequently the spatial
average in (\ref{d.matrix}). We will study in more detail
this situation later in the case of an the collective emission
properties of atomic ensambles.

\section{Atom-Photon mapping in a square lattice}
\label{section.lattice}
The situation in which atoms are arranged in a crystal is found in
experimental setups such as ultracold atoms in optical lattices
and Coulomb crystals of trapped ions.
The results presented in the previous section are applied here to
study the collective emission process in these systems.
We obtain analytical results by using the plane-wave approximation.

\subsection{General Discussion}

Let us study for concreteness the case of one (1D) or three dimensional (3D) square lattices, although our results are easily generalized to different lattice geometries.
Assuming periodic boundary conditions, the allowed
wavevectors are
\begin{equation}
\KK_\nn = \frac{2 \pi}{d_0}
\sum_\alpha \frac{n_\alpha}{N_\alpha} \hat{\alpha}.
\label{pbc.wavevectors}
\end{equation}
In the 1D case we consider that the atom chain is aligned in the $z$
direction. Thus, in (\ref{pbc.wavevectors}) and the forecoming
expressions, $\alpha$ runs over
$\alpha = z$ in 1D, and $\alpha = x, y, z$  in 3D.
$\hat{\alpha}$ is a unit vector in the direction $\alpha$, and
$N_\alpha$ is the number of atoms along $\hat{\alpha}$.
Considering, for concreteness, the case of even $N_\alpha$, 
each wavevector is determined by the set of integers
$n_\alpha = - N_\alpha/2, \dots, N_\alpha/2 - 1$.
By applying Eq. (\ref{photon.distribution.collective}) we determine
the photon number angular distribution for the photonic state emitted by a spin-wave
excitation with momentum $\KK_\nn$,
\begin{eqnarray}
&& \hspace{-1.cm} I_\nn(\Omega) =
\frac{1}{4 \pi}
\frac{\bar{\Gamma}}{\Gamma_{\bf \nn}}
\label{semiclassical.crystal}
\\
& & 
\hspace{0.5cm}
\frac{1}{N}
\prod_{\alpha}
\frac{\sin^2( ( k_\LL \uu_\Omega^\alpha - \kk_\LL^\alpha - \KK_\nn^\alpha)
  d_0 N_\alpha / 2)}
     {\sin^2( ( k_\LL \uu_\Omega^\alpha - \kk_\LL^\alpha -
       \KK_\nn^\alpha) d_0 /2)} \nonumber .
\end{eqnarray}
$I_\nn(\Omega)$ shows a series of
diffraction maxima at solid angles $\Omega$ at which the $\sin^2$-function in the
denominator vanishes. 
Note that by including the photon polarization, we would have got an
additional function of $\Omega$ multiplying the photon distribution $I_n(\Omega)$,
which would correspond to the single atom dipole pattern. 
The latter would induce the suppression of diffraction peaks, if they are in
a direction forbidden by the dipole pattern.
Since we are specifically interested on collective effects we do not consider this effect
in the discussion that follows.

To get a quantitative description of the photon distribution
we notice first that
\begin{equation}
\frac{1}{N}
\frac{\sin^2(N x)}{\sin^2(x)} \approx \sum_m f_N(x - m \pi)
\hspace{0.5cm} \textmd{if} \hspace{0.2cm} \left( N \gg 1 \right) ,
\label{sinc.approx}
\end{equation}
where the function $f_N(x)$ describes the shape of each of the diffraction peaks,
\begin{eqnarray}
f_N(x) &=& N {\rm sinc}^2(N x), \ \  -\pi/2 < x < \pi/2 , \nonumber \\
f_N(x) &=& 0, \ \ \ \ {\rm otherwise}.
\label{define.f}
\end{eqnarray}
Thus in the limit $N_\alpha  \gg 1$,
one can approximate the emission probability as a sum over
Bragg scattering contributions,
\begin{eqnarray}
I_\nn (\Omega) &=& \sum_{\bf m} I^{[{\bf m}]}_\nn(\Omega) ,
\nonumber \\
I_\nn^{[{\bf m}]} (\Omega)
&=& \frac{1}{4 \pi} \frac{\bar{\Gamma}}{\Gamma_{\nn}}
\label{sinc.approximation} \\
& & \prod_\alpha
f_{N_\alpha}\left(
\frac{k_\LL d_0 \nn^\alpha_\Omega}{2}  -
\frac{\kk^\alpha_\LL d_0}{2} -
\frac{n_\alpha}{N_\alpha}  \pi
+ m_\alpha  \pi \right) . \nonumber
\end{eqnarray}
Each term in the sum is labeled by the vector ${\bf m}$, and
corresponds to a different diffraction peak.
The probability that the spin-wave $\nn$ emits a photon in the $\mm$ diffraction peak is given by
\begin{equation}
p_\nn^{[\mm]} = \int d \Omega \ I^{[\mm]}_\nn (\Omega) .
\label{def.pm}
\end{equation}

Eq. (\ref{sinc.approximation}) has
a clear interpretation in terms of momentum conservation.
$I_\nn(\Omega)$ has a maximum whenever there is a value of $\mm$ such
that

\begin{equation}
k_\LL \uu^\alpha_\Omega
= \kk_\LL^\alpha + n_\alpha \frac{2 \pi}{d_0 N_\alpha} + m_\alpha
\frac{2 \pi}{d_0}, \ \ \forall \ \alpha.
\label{momentum.conservation}
\end{equation}
That is, the linear momentum of the emitted photon has to match
the sum of three contributions: the momentum of the incident laser,
$\kk_\LL$; the initial momentum of the spin-wave, $\nn 2 \pi/(d_0
N_\alpha)$, and the contribution from the lattice periodicity, which
enters through the reciprocal wavector $\mm 2 \pi/d_0$. In 1D,
condition (\ref{momentum.conservation}) has to be satisfied only by
the $\alpha = z$ component of these vectors, that is, only the projection of the
momentum on the chain is conserved. In 3D, on the contrary, the equality has to
be satisfied by all the vector components. The relation
(\ref{momentum.conservation}) determines the maxima in the emission
pattern depending on $d_0$ and $\lambda$, but it also determines the
collective rates, through the normalization condition on
$I_\nn(\Omega)$, see Eq. (\ref{collective.rates}).

Since $-1 \leq \uu^\alpha_\Omega \leq 1$, if $\lambda > 2 d_0$
there is at most a single value of $\mm$, for which condition
(\ref{momentum.conservation}) can be fulfilled for some vector
$\uu_\Omega$. 
Thus, $\lambda > 2 d_0$ is the directional regime in the emission of
photons, whereas in the case $\lambda < 2 d_0$, we cannot ensure that
photons are collimated in a single direction. In the following two
subsections, we will study these regimes in 1D and 3D. 
In particular, we will be interested in determining $\Delta
\theta$, the angular width of the photon beam in the directional
regime, and $\Gamma_\nn$, the collective emission rates. These
quantities will be studied as a function of dimensionality, $d_0$,
$\lambda$, and $N$. 

\subsection{Atom chains}
\begin{figure}[h]
\center
\resizebox{!}{!}{\includegraphics{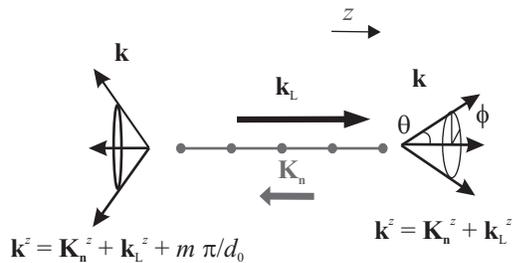}}
\caption{Photon emission by a spin-wave with linear momentum $\KK_n$
  in an atomic chain. The maxima in the photon distribution are at
  angles $\theta$ such that momentum along the chain is conserved, up to
  reciprocal lattice wave vectors.}
\label{fig.momentum.conservation.1D}
\end{figure}

%
%
We consider for concreteness that $\kk_\LL$ points in the $z$
direction, parallel to the chain axis. 
The vectors $\nn$, $\mm$ are reduced now to scalars $n$, $m$, that
correspond to the projections on the chain axis.
Due to the symmetry of the
problem, the photon distribution emitted by a spin-wave $n$
depends only on the angle $\theta$, through $\uu^z_\Omega = \cos \theta$, 
\begin{eqnarray}
& & \hspace{-1.cm} I_{n}(\Omega) = \nonumber \\
& &
\frac{1}{4 \pi}
\frac{\bar{\Gamma}}{\Gamma_n}
\frac{1}{N}
\frac{\sin^2 \left( ( k_\LL d_0 (\uu^z_\Omega - 1) - n \frac{2 \pi}{N}  ) N / 2\right)}
{\sin^2 \left( (k_\LL d_0 (\uu^z_\Omega - 1)  - n \frac{2 \pi}{N} )/2 \right)}.
\label{collective.1D.0}
\end{eqnarray}
Taking the limit $N \gg 1$ and using (\ref{sinc.approx}),

\begin{eqnarray}
I_n (\Omega) &=& \sum_m I^{[m]}_n(\Omega) ,
\label{sinc.approximation.1D} \\
I_n^{[m]} (\Omega)
&=& \frac{1}{4 \pi} \frac{\bar{\Gamma}}{\Gamma_n}
f_N
\left(
\frac{k_\LL d_0}{2} (\uu^z_\Omega - 1) - n \frac{\pi}{N} + m \pi \right) . \nonumber
\label{photon.distribution.1D}
\end{eqnarray}
The peaks in the photon distribution correspond to the emission of photons such
that the $z$ component of momentum is conserved,
\begin{equation}
\uu^z_\Omega - 1   = \frac{\lambda}{d_0} \frac{n}{N} - \frac{\lambda}{d_0} m,
\label{momentum.conservation.1D}
\end{equation}
that is, there is a maximum whenever there is a value of $m$ which satisfies this
relation. 
In the directional regime, $\lambda > 2 d_0$, there is at
most a single value of $m$ which satisfies (\ref{momentum.conservation.1D}). 
However, in this case, momentum conservation only determines the
value of $\theta$ at the emission maximum, which implies that, in general,
photons are emitted in cones spanned by different values of $\phi$. 
Only when the maximum happens at $\theta
= 0$, or $\theta = \pi$, photons are collimated in the forward- or
backward-scattering directions, respectively (see Fig. \ref{fig.momentum.conservation.1D}).
%
%

\begin{figure}[h]
\center
\resizebox{!}{!}{\includegraphics{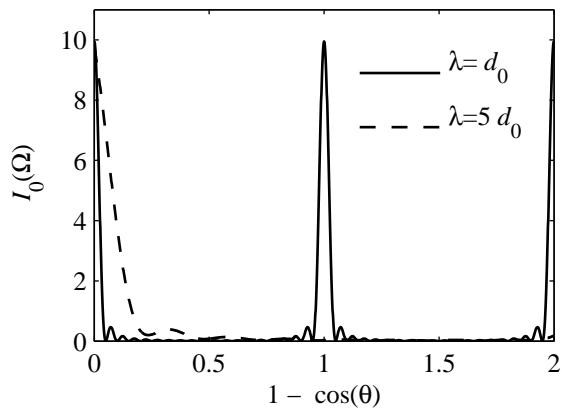}}
\caption{Angular photon distribution of the photonic mode emitted by
  the completly symmetric spin-wave in a chain of
$N = 20$ atoms. We have used the exact egenvalues of the density
matrix, and followed the method described in the previous section.}
\label{fig.ang1}
\end{figure}

\begin{figure}[h]
\center
\resizebox{!}{!}{\includegraphics{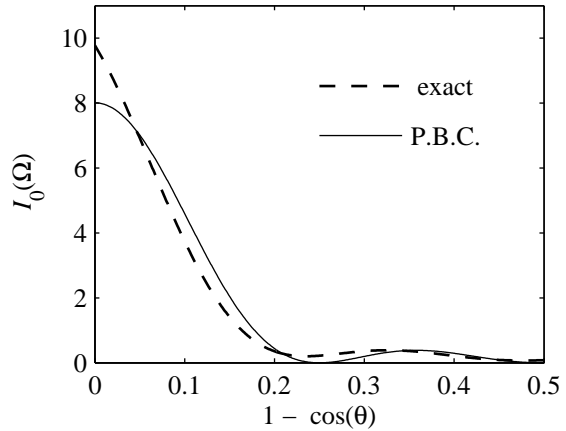}}
\caption{Angular photon distribution of the photonic mode emitted by
  the completly symmetric spin-wave from a chain of $N = 20$ atoms,
  $\lambda > 5 d_0$,
  calculated assuming periodic boundary conditions  
  (continuous line), and performing and exact
  diagonalization of the master equation (dashed line). The
  distribution in the exact case shows a departure form the
  ${\rm sinc}^2$-shape due to collective effects.}
\label{fig.ang2}
\end{figure}

Let us study first the case $n = 0$, that is, the emission
properties of the completely symmetric state.
The forward-scattering contribution, $I_0^{[0]}$, 
has an angular width given by
$\Delta \theta_{1D} = 1/\sqrt{k_\LL d_0 N}$.
The contribution to the emission pattern from each of the
Bragg terms is
\begin{eqnarray}
p^{[0]}_0  &=& \int d \Omega  I_0^{[0]} (\Omega) =
\frac{\bar{\Gamma}}{\Gamma_0} \frac{\lambda}{4 d_0},
\nonumber \\
p^{[m]}_0 &=& 2 p^{[0]}_0, \hspace{0.5 cm} m \neq 0.
\end{eqnarray}
Condition (\ref{momentum.conservation.1D}) leads to the result that
the number of emission cones with $m \neq 0$ is given by $\iint (2 d_0 /
\lambda)$. A calculation made without resorting to the planewave
approximation, yields the same results, even for relatively small atom
numbers ($N = 20$), see Fig. \ref{fig.ang1}. 
However, the shape of the photon angular distribution in the exact
calculation  shows a
departure from the ${\rm sinc}^2$-shape predicted by
Eq. (\ref{photon.distribution.1D}), see Fig. \ref{fig.ang2}. 
By using the normalization
condition for $I_0(\Omega)$, we determine the probability of
emission in the forward-scattering direction,
\begin{equation}
p_0^{[0]} = 
\frac{1}{1 + 2 \ \iint\left( \frac{2 d_0}{\lambda} \right)}.
\label{fidelity.1D}
\end{equation}
Thus, in the directional regime, $\lambda > 2 d_0$, all the diffraction peaks but
the forward-scattering one are suppressed. The collective emission
rate is given by
\begin{eqnarray}
\frac{\Gamma_{0}}{\bar{\Gamma}}
&=&
\frac{\lambda}{4 d_0}
+ \frac{\lambda}{2 d_0} {\rm int} \left( \frac{2 d_0}{\lambda} \right) .
\label{collective.rate1D}
\end{eqnarray}
%
%

\begin{figure}[h]
\center
\resizebox{!}{!}{\includegraphics{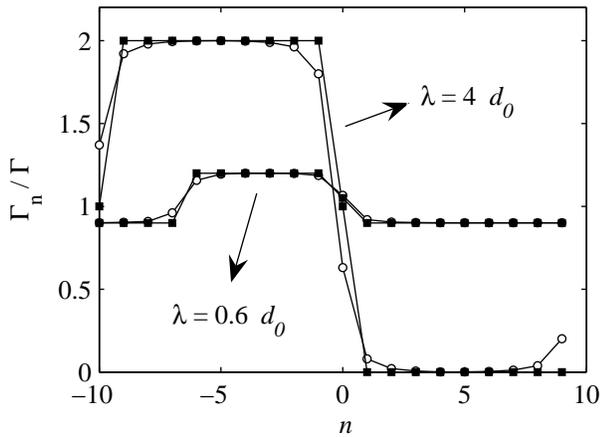}}
\caption{Calculation of the collective rates of a chain with $N
  = 20$ atoms. Filled squares: result predicted by the planewave
  approximation (\ref{rates.1D}). Empty circles: eigenvalues of the
  matrix $J_{ij}$ defined by (\ref{master.eq.coeff}). In the last case, to
  each eigenvalue of $J_{ij}$ we assign a wavenumber $n$ that corresponds
  to the maximum Fourier component of the corresponding eigenvector.}
\label{fig.rates}
\end{figure}

Finally, we calculate the emission rate for all the 1D spin-wave
states. 
They can be written as an integration over the contributions 
coming from different diffraction peaks, through the normalization
condition, we have
\begin{equation}
\frac{\Gamma_n}{\bar{\Gamma}} =
\frac{1}{4 \pi}
\sum_m
\int_{-1}^1
d x
f_N
\left(
\frac{k_\LL d_0}{2} (x - 1) - n \frac{\pi}{N} + m \pi \right).
\label{normalization.1D}
\end{equation}
The collective rates are thus determined by the number of diffraction
peaks which appear in the emission pattern of each collective state
$n$. Considering the limit in which $k_\LL d_0 N \gg 1$, that is, the
delta limit for each of these peaks, 
we get the following distribution of emission rates,
\begin{eqnarray}
&& \hspace{-0.5cm} 
\frac{\Gamma_n}{\bar{\Gamma}} =
\nonumber \\
&&
\hspace{-0.25cm}
\chi_{1D}
\left((
\theta (-\frac{n}{N} ) \theta ( \frac{2 d_0}{\lambda}
  + \frac{n}{N} ) +
\iint ( \frac{2 d_0}{\lambda} + \frac{n}{N} )
\right) , 
\label{rates.1D}
\end{eqnarray}
with $\chi_{1D} = \lambda/(2 d_0)$, and $\theta(x)$ is the Heaviside function.
This expression describes quite well the emission rates calculated
without assuming periodic boundary conditions, that is, 
by diagonalizing the matrix $J_{ij}$
given by (\ref{master.eq.coeff}), see Fig. \ref{fig.rates}.
In the limit $d_0 \gg \lambda$, we recover $\Gamma_n =
\bar{\Gamma}$. 
On the contrary if $\lambda$ is comparable to $d_0$, emission rates
are renormalized.
In the directional regime,
$\lambda > 2 d_0$, some of the states have an enhanced rate,
$\Gamma_n = \chi_{1D}$ (superradiant), whereas there are states for which
$\Gamma_n = 0$ (subradiant) \cite{note.finite}. 
Due to the relation between the
photon distribution and the emission rates (\ref{normalization.1D}),
subradiant states correspond to spin-waves whose emission pattern is
not peaked at a given value of $\theta$. On the contrary, superradiant
states are spin-waves whose emission pattern does contain a maximum as a
function of $\theta$. Note that Eq. (\ref{rates.1D}) does not describe the case of Dicke
superradiance, which would predict a single superradiant state, with
$\Gamma_0 = N \bar{\Gamma}$. The reason is that we have assumed
condition $N k_\LL d_0 \gg 1$, that is, where are always in the regime
in which the light wavelength is much smaller than the size of the
chain.
Our results for 1D are consistent with Ref. \cite{Carmichael}, where
the collective light emission form an atomic chain was studied with the
quantum jump formalism.

%
%
%

To summarize the situation in 1D, in the directional regime $\lambda > 2 d_0$,
superradiant spin-waves emit photons into a single emission cone, with
a rate that is enhanced by $\chi_{1D}$. 
This effect can be used to
generate photons that are collimated along the chain axis. Note that,
in general, the atom-photon mapping induced with the $\Lambda$-scheme
of Fig. \ref{fig.lambda} may compete with other radiative processes,
such as the radiative decay from $| a \rangle$ back to $| s \rangle$, or
the radiative decay from $| a \rangle$ to other atomic levels that are not
included in the $\Lambda$-scheme. This problem can be solved by
enhancing the atom-photon mapping rate, choosing $\chi_{1D} \gg
1$. Since $\chi_{1D}$ does not depend on the chain size, this implies
to choose $\lambda \gg d_0$. Another way out of this problem is to use
a cycling transition (see Appendix \ref{appendix.cycling}).

\subsection{3D Atom lattices}

Contrary to the 1D case, in 3D it is not simple to obtain closed
expressions to describe the emission in the planewave
approximation.
In order to get a simpler picture, we replace the ${\rm
  sinc}$-function 
in the definition (\ref{sinc.approx}) by a gaussian
which is normalized in the same way,
\begin{equation}
f_N(x) \approx \sqrt{\pi} N_x e^{- x^2 N^2} .
\end{equation}
This approximation is justified in the limit in which the diffraction
peaks are narrow enough, such that they do not overlap, that is,
$N_\alpha k_\LL d_0 \gg 1$.
Assuming that the length of the lattice is the same in any spatial
direction, $N_x = N_y = N_z$, then
the angular photon distribution
for a state ${\bf n}$ is given by
\begin{eqnarray}
I_\nn(\Omega) &=& \sum_\mm I_\nn^{[\mm]}(\Omega),
\label{angular.3D} \\
I_\nn^{[\mm]}(\Omega) &=&
\frac{1}{4 \pi} \frac{\bar{\Gamma}}{\Gamma_\nn}
\pi^{3/2} N
e^{-
\left( \frac{k_\LL d_0}{2} \uu_\Omega - \frac{\kk_\LL d_0}{2}
- \pi \frac{\nn}{N} - \pi \mm \right)^2 N_x^2}.
\nonumber
\end{eqnarray}
A given term
$I^{[\mm]}_\nn(\Omega)$, has a non-negligible contribution only if
\begin{equation}
\frac{|\kk_\LL d_0 + 2 \pi \nn / N + 2 \pi \mm|}{k_\LL d_0} = 1 ,
\label{condition.3D}
\end{equation}
since otherwise, there are no photons which satisfy the
energy-momentum conservation.

Let us study first the directional regime, $\lambda > 2 d_0$. In this
case we find the following two possible situations:

(i) Superradiant states -- Spin-waves for which there exists a value of $\mm$, say $\mm_c$,
such that (\ref{condition.3D}) is satisfied. The condition $\lambda >
2 d_0$ ensures that there is a single value $\mm_c$, such that the
photon distribution can be simplified,
\begin{equation}
I_\nn(\Omega) = I_\nn^{[\mm_c]}(\Omega).
\label{I.mc}
\end{equation}  
Thus, there is a single emission peak in the direction,
\begin{equation}
\uu_{\Omega_{\rm max}} = \frac{\kk_\LL d_0 + 2 \pi \nn/N + 2 \pi
  \mm_c}{k_\LL d_0}.
\label{maximum.3D}
\end{equation}
Photons are collimated in a beam with width $\Delta \theta_{3D} = 1/(k_\LL d_0 N^{1/3})$.
From (\ref{I.mc}), and the normalization condition, we can determine
the emission rate,
\begin{eqnarray}
\frac{\Gamma_\nn}{\bar{\Gamma}} &=&
\nonumber \\
& & \hspace{-0.5cm}
\frac{N}{4 \pi} \int d \Omega \ \pi^{3/2}
e^{- \left( \frac{k_\LL d_0}{2} \nn_\Omega
- \frac{\kk_\LL d_0}{2} - \pi
    \frac{\nn}{N} - \pi \mm_c\right)^2
  N_x^2 } =
\nonumber \\
& & \hspace{-0.5cm}
\frac{1}{\sqrt{\pi}} N_x \left( \frac{\lambda}{2 d_0}\right)^2
= \chi_{3D}.
\label{rate.3D.1}
\end{eqnarray}
Thus, superradiant states have a decay rate that is enhanced by the
optical thickness, $\chi_{3D}$.  

(ii) Subradiant states -- Spin-waves for which there is no value of
$\mm$ which satisfies (\ref{condition.3D}). In this case, there is not
a single dominant contribution $\mm$, such that,
directionality in the emission of photons is not guaranteed. Also, the
normalization condition on $I(\Omega)$ leads directly to the result
$\Gamma_\nn = 0$ \cite{note.finite}.

From this analysis, we conclude that superradiant spin-waves are
interesting in the context of quantum information, since they emit
collimated photons with an enhanced emission rate.
However the emission direction
is determined by a value of $\mm_c$ which has to be calculated for each
particular spin-wave. There are two relevant cases in which this situation
becomes simpler, since energy-momentum conservation ensures that the
only Bragg
contribution is $\mm_{\rm c} = \00$. The first one is the case $\lambda > 4
d_0$, because
\begin{eqnarray}
|k_\LL \uu_\Omega^\alpha - \kk_\LL^\alpha| &<& 2 k_\LL, 
\nonumber \\
|\frac{n_\alpha}{N} \frac{2 \pi}{d_0} + m_\alpha \frac{2 \pi}{d_0} |
&>& \frac{1}{2} \frac{2 \pi}{d_0} \ \ ({\rm if} \ \ \mm \neq \00),
\label{condition.fs3D}
\end{eqnarray}
and thus for values $\lambda > 4 d_0$
the energy-momentum conservation condition
 (\ref{momentum.conservation}) is only fulfilled by
$\mm_{\rm c} = \00$. Second, assume that the spin-wave is created by the absorption
of a photon within the $\Lambda$-configuration of
Fig. \ref{fig.lambda}, such that the spin-wave linear momentum can be
written like $\KK_\nn = \qq - \kk_\LL$, with $q = k_\LL$. Then, the
energy-momentum condition is now
\begin{equation}
\qq + \kk = \mm \frac{2 \pi}{d_0},
\label{condition.fs3D.2}
\end{equation}
which is satisfied only by $\mm_{\rm c} = \00$ if $\lambda > 2 d_0$. Indeed,
the spin-wave created in this way is always superradiant, since
$\KK_\nn$ automatically satisfies Eq. (\ref{condition.3D}).
A similar result is obtained in 
Ref. \cite{Scully08} for the case of a cloud of atoms, where the rate
of spontaneous emission of a state created by the absorption of a
photon, is shown to increase with the optical thickness.

To summarize the situation in 3D, in the directional regime $\lambda >
2 d_0$, superradiant spin-waves emit collimated photons with a
rate enhanced by $\chi_{3D}$. Contrary to the situation found in 1D,
in 3D lattices $\chi_{3D}\propto N_x$, and thus it can be increased by
increasing the size of the system. In this way, the atom-photon
mapping could be much faster than other competing decay channels. 

\section{Emission of light by hot atomic ensembles}
\label{section.ensembles}
The formalism presented in the previous section
can be extended to situations in which atoms
are not at fixed positions in space. In this case, one expects the
photonic state not to be a pure state, but a mixed one.
This is relevant to the description of hot atomic vapors.
We follow the discussion presented in subsection \ref{effects.motion},
and consider the case of fast motion ($\tau_{\rm mot} \ll \Gamma_n$).
The results presented here are in agreement with Ref. \cite{duan.pra},
where the photonic mode emitted by an atomic ensemble is studied.
Our first task is to reformulate the problem of photon emission by
finding the master equation that describes this situation. 
\subsection{Master Equation for atomic ensembles}
In the fast motion limit,
particle positions may be described as a set of
independent random variables,
$\{ \rr_j  \}$. We choose for simplicity a gaussian distribution probability,
\begin{equation}
\rho(\rr) = \frac{1}{\pi^{3/2} L^3} e^{- (r/L)^2} .
\label{probability.distribution}
\end{equation}
The particle positions fulfill the following identity, which will turn
out to be the basis for the following calculations,
\begin{equation}
\langle e^{-i \qq (\rr_j - \rr_l)} \rangle_{\rm mot} = \delta_{j l} +
\left( 1 - \delta_{j l} \right) e^{- q^2 L^2 / 2} ,
\label{exponential.average}
\end{equation}
where $\langle \dots \rangle_{\rm mot}$ is an average over the atomic
positions.

To calculate the coefficients of the master equation we start from
Eq. (\ref{master.eq.coeff.0}) and perform the following average,
\begin{equation}
J_{ij}
=
\pi \sum_\kk g_k^2 \
\delta(\omega_k - \omega_\LL)
\langle e^{i ({\bf k} - {\bf k}_\LL)(\rr_i - \rr_j)} \rangle_{\rm mot},
\label{master.eq.coeff.average}
\end{equation}
were we have neglected the Cauchy principal value contribution, since
it leads to an energy shift that does not play any role in the discussion
that follows. After using (\ref{exponential.average}), integrating
over $\kk$, and taking the limit $k_\LL L \gg 1$, we get
\begin{eqnarray}
J_{ii} &=& \bar{\Gamma}/2, \nonumber \\
J_{ij} &=& \frac{\bar{\Gamma}}{4 (k_\LL L)^2} .
\hspace{0.2cm} \ \ (i \neq j) .
\label{master.eq.coeff.ensembles}
\end{eqnarray}
One can readly diagonalize the matrix $J_{ij}$ and obtain the
eigenspaces of the master equation. The first is the completely
symmetric state ($n = 0$),
\begin{equation}
{\cal M}_{j 0} = \frac{1}{\sqrt{N}}, \hspace{0.2cm}
J_0 = \frac{\bar{\Gamma}}{2} \left(\chi_{\rm en} + 1 \right) ,
\end{equation}
where $\chi_{\rm en} = (N-1)/(2(k_\LL L)^2)$ is the optical thickness
of the atomic ensemble. Note that in the limit $N \gg 1$, and defining
the atom density, $n_{\aat} = N/L^3$, the optical thickness can be
recast in the more familiar form
$\chi_{\rm en}= n_{\aat} \lambdabar^2 L / 2 $.
The second eigenspace is spanned by the spin-waves orthogonal to ${\cal
M}_{j 0}$,
\begin{equation}
\sum_j {\cal M}_{jn} = 0, \hspace{0.2cm} 
J_n = \frac{\bar{\Gamma}}{2} \left( 1 - \frac{1}{2 (k_\LL L)^2}
\right), \hspace{0.2cm}  {\rm if} \ n \neq 0.
\end{equation}
Collective effects happen if $\chi_{\rm en}
\gg 1$. In this case, there is a single superradiant spin-wave mode,
corresponding to the completely symmetric state, and $N-1$ states
which decay with the single atom emission rate, $J_n = \bar{\Gamma}/2$.
\subsection{Single photon state}
The spin-wave mode $n=0$ is the only one
to show collective effects, and it is the collective
state which can be created in experiments with atomic vapors. For
these reasons we focus on the following on its properties. First, we
consider the emission of light by the initial atomic state with a
single excitation,
\begin{equation}
| \Psi_0 \rangle_\aat = \frac{1}{\sqrt{N}} \sum_j b_j^+ | 0 \rangle_\aat .
\end{equation}
The atom-photon mapping can be extended to this situation by
considering first that the atomic state is mapped into a given
photonic state, and then by performing the average on the atomic
positions. This gives as a result a photon density matrix,
\begin{eqnarray}
| \Psi_0 \rangle_\aat | 0 \rangle_\pph \langle 0 | &\to& | 0 \rangle_\aat \rho,
\nonumber \\
\rho &=& \sum_{\kk,\kk'} \rho_{\kk \kk'}
a^+_\kk | 0 \rangle_\pph \langle 0 | a_{\kk'} .
\end{eqnarray}
Note that the mapping is now form pure to mixed states. Unfortunately,
with the statistical properties of the atomic positions considered
here, it is not possible to ensure that this is still a gaussian
map. We can, however, use Eqs. (\ref{mapping.pbc},\ref{exponential.average}) to get
\begin{eqnarray}
\rho_{\kk \kk'} &=&
\langle \phi_{0, \kk} \phi_{0, \kk'}^*  \rangle_{\rm mot} =
(1 - \epsilon) \bar{\phi}_{0,\kk} { \bar{\phi}_{0,\kk'} }^*
+ \epsilon \bar{\rho} , \nonumber \\
\epsilon &=& \frac{1}{1 + \chi_{\rm en}}.
\label{photon.desnsity.matrix}
\end{eqnarray}
We work, for clarity, in the interaction picture with respect to
$H_{\rm lm}$. The coherent component of the photon
density matrix is the pure state $| \bar{\phi}_0 \rangle_\pph$,
\begin{equation}
\bar{\phi}_{0,\kk} = \sqrt{1 + \frac{1}{\chi_{\rm en}}} \sqrt{N-1} g_k \frac{e^{-|\kk - \kk_\LL|^2
  L^2 / 4}}{-i(\omega_k - \omega_\LL) + J_0},
\label{pure.state}
\end{equation}
whereas $\bar{\rho}$ is a normalized mixed state which describes
the incoherent (isotropic) component,
\begin{eqnarray}
\bar{\rho}_{\kk \kk'} =
\frac{(1 + \chi_{\rm en}) g_k g_{k'}
e^{-|\kk - \kk'|^2 L^2/4}}{\left(-i(\omega_k - \omega_\LL) + J_0 \right)
                    \left(i(\omega_{k'} - \omega_\LL) +
                      J_0 \right)}.
\end{eqnarray}
Under the condition $\chi_{\rm en}
\gg 1$, the main contribution to the photon density matrix is the
pure one, $| \phi_{0} \rangle_\pph$. The parameters which describe
the directionality in the photon emission can be readly
evaluated. The emission pattern has both coherent and incoherent
contributions,
\begin{eqnarray}
I_0(\Omega) &=& I_0^{\rm coh}(\Omega) + I_0^{\rm inc}(\Omega),
 \nonumber \\
I_0^{\rm coh}(\Omega) &=&
(1 - \epsilon)
\frac{N - 1}{4 \pi \chi_{\rm en}} e^{-(k_\LL \uu_\Omega - \kk_\LL)^2 L^2 /
2},
\nonumber \\
I_0^{\rm inc}(\Omega) &=& \epsilon \frac{1}{4 \pi} .
\end{eqnarray}
Thus, photons are collimated in the forward-scattering direction,
with angular width $\Delta
\theta_{\rm en} = 1/(k_\LL L)$. The deviation of the photon emission
from directionality is determined by the probability of emission out
of the forward-scattering cone,
${\cal E} = \int d \Omega I_0^{\rm inc}(\Omega) = \epsilon$. 

Finally, we calculate the purity of the photon state which is defined
as
\begin{eqnarray}
{\cal P} &=&  {\rm Tr}(\rho^2) \nonumber \\
&=&  (1 - \epsilon)^2 + \epsilon ^2
{\rm Tr}\left(\bar{\rho}^2 \right)
+ 2 \epsilon (1-\epsilon) \langle \bar{\phi}_0 | \bar{\rho } | \bar{\phi}_0 \rangle .
\end{eqnarray}
This quantity is very relevant when using photons in quantum
information processing, since it describes the efficiency in the
process of interference of photonic modes in a beam-splitter, being
${\cal P} = 1$, the case corresponding to a pure state.
To calculate ${\cal P}$, we notice first that
\begin{equation}
\TTr \left( \bar{\rho}^2 \right) = \sum_{\kk,\kk'} |\rho_{\kk,
  \kk'}|^2 = \frac{1}{2(k_\LL L)^2} .
\end{equation}
Also because of the Cauchy-Schwartz inequality, this implies that
\begin{equation}
|\langle \phi_0 | \bar{\rho} | \phi_0 \rangle|^2
\leq \TTr \left( \bar{\rho}^2 \right),
\end{equation}
such that,
\begin{equation}
{\cal P} = {\rm Tr}(\rho^2) = (1 - \epsilon)^2 + {\cal O}( 1/ k_\LL L)
\approx 1 - 2/\chi_{\rm en} .
\end{equation}
Thus, in the limit $k_\LL L \gg 1$, the purity of the photon state is
solely determined by $\chi_{3D}$.

\subsection{Multiphoton state}
We study now the properties of the multiphoton case.
The mapping is from a state with $M$ atomic excitations 
in the completely symmetric state to a multiphoton mixed
state,
\begin{equation}
\frac{1}{\sqrt{M !}}
\left( \frac{1}{\sqrt{N}} \sum_j b^+_j \right)^M
| 0 \rangle_\aat | 0 \rangle_\pph \to | 0 \rangle_\aat \rho.
\label{multi.photon.mapping}
\end{equation}
We assume the low-excitation limit ($M \ll N$). 
In Appendix \ref{appendix.multi.photon}, we show that the mixed photonic state can be written
like 
\begin{eqnarray}
\rho  &=& (1 - \epsilon)^M | \Phi \rangle_\pph \langle \Phi| +  \tilde{\rho},
\nonumber \\
|\Phi \rangle_\pph &=& \frac{1}{\sqrt{M !}}
\left( \sum_\kk \bar{\phi}_{0,\kk} a^+_\kk \right)^M | 0 \rangle_\pph .
\label{multi.photon.dm}
\end{eqnarray}
That is, it is a sum of a pure state consisting of $M$ photons in the
photonic mode defined by (\ref{pure.state}), and the (not normalized)
mixed state $\tilde{\rho}$. In order to determine the
purity of $\rho^{\pph}$ we should study the mixed contribution, which
is far more complicated than in the single photon case. Insted, we
notice that
\begin{equation}
{\cal P} = 
\TTr \left( \rho^2 \right) \geq (1 - \epsilon)^{2 M},
\label{multi.photon.purity}
\end{equation} 
which allows us to obtain a lower bound for the purity.

\section{Implementations}
\subsection{Collective light emission in ultracold atoms in optical lattices}
Ultracold atoms in optical lattices are an ideal system for the
observation of the effects described in this work. For example, atoms
in a Mott phase \cite{Greiner} would be ideally suited to study collective light
emission from a square lattice.
In this setup,
atoms are placed at distances that are comparable to
optical wavelengths, since potential wells in a
standing--wave are indeed separated by
$d_0 = \lambda_{\textmd{sw}}/2$, with $\lambda_{\textmd{sw}}$, the
wavelength of the counterpropagating lasers that create the lattice.
The conditions for the directional regime are met by using an optical
transition such that $\lambda > \lambda_{\textmd{sw}}$.  Under this
conditions, ultracold atoms in optical lattices are ideal to form an
atom-light quantum interface \cite{quantum.memories,Hammerer,DuanCiracZoller}, 
where the quantum state of light can be indeed manipulated, as show
recently in \cite{Muschik}. The properties of the emitted light may
also be used to measure properties of quantum many-body phases in
optical lattices \cite{deVega.08}. Our theory can also be applied to
the collective matter-wave emission in optical lattices proposed in
Ref. \cite{deVega.arX}.

We propose now an experiment in which the renormalization of
the collective rates may be observed by performing the following
steps (see Fig. \ref{fig.optical.lattice}):

\begin{figure}[h]
\center
\resizebox{!}{!}{\includegraphics{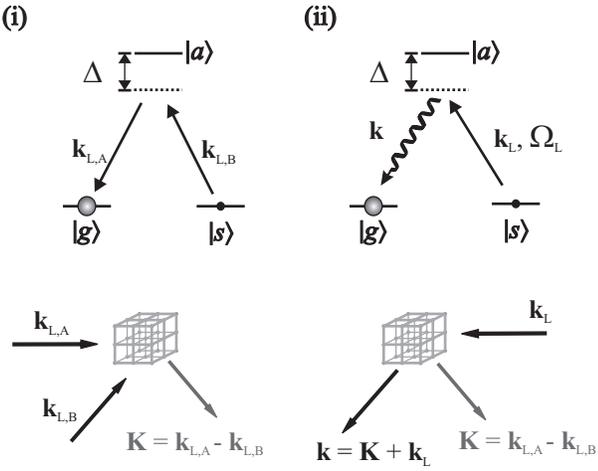}}
\caption{(i) Creation of a spin-wave with linear momentum $\kk_{L,A} +
  \kk_{L,B}$ in an optical lattice.
(ii) Mapping from the spin-wave to a photonic mode.}
\label{fig.optical.lattice}
\end{figure}
(i) First, we use a protocol to initialize the atomic quantum state of the
lattice. Assume that all the atoms are initially in the ground state
$| g \rangle$. The $\Lambda$-scheme of Fig. \ref{fig.lambda} may be used for
the initialization by shining the lattice during a short time $T$ with two
lasers with wavevectors $\kk_{\LL,\AA}$, $\kk_{\LL, \BB}$,
and Rabi frequencies $\Omega_{\LL, \AA}$, $\Omega_{\LL, \BB}$, and
detuning $\Delta$, such that $\Omega_{\LL, \AA}, \Omega_{\LL, \BB} \ll
\Delta$.
Under these conditions, a coherent state of
spin-waves with momentum $\KK = \kk_{\LL,\AA} - \kk_{\LL,\BB}$ is created,
\begin{eqnarray}
| \Psi \rangle_\aat &\propto& e^{-i (\Omega_{\rm AB} T / 2) b^+_K} | 0 \rangle_\aat ,
\nonumber \\
b^+_\KK &=& \frac{1}{\sqrt{N}} \sum_j e^{i \KK \rr_j} b^+_j ,
\end{eqnarray}
where $\Omega_{\rm AB} = \Omega_{\LL,\AA} \Omega_{\LL,\BB}/\Delta$,
and condition $\Omega_{\rm AB} T \ll 1$ has to hold in order to ensure
that the system is in the low excitation regime.

(ii) After the creation of the collective atomic state, this can be realeased by
a second laser with momentum $\kk_{\LL}$ in the $\Lambda$-scheme considered
along this work. The analysis presented in section
\ref{section.lattice} can be applied to study the emission of photons
by setting $\KK_\nn = \kk_{\LL,\AA} - \kk_{\LL,\BB}$.
For example, if $\kk_{\LL} = \kk_{\LL,\BB}$, then the spin-wave
emits a photon in the direction $\kk_{\LL,\AA}$, with a superradiant rate
$\chi_{3D} \bar{\Gamma}$. On the contrary, if $\KK_\nn$ does not
satisfy the condition (\ref{condition.3D}), then the emission rate
will be suppressed, up to finite size effects.

\subsection{Deterministic generation of entangled states of photons}

An application of our ideas to generate photons in a deterministic way
requires a system where experimentalists are both able to reach the
directional regime of collective light emission, and to initialize the
atomic system in a given spin-wave state. This idea may find useful
applications in quantum cryptography \cite{Gisin}, quantum computation
\cite{KLM}, and quantum litography \cite{Boto}.

\begin{figure}[h]
\center
\resizebox{!}{!}{\includegraphics{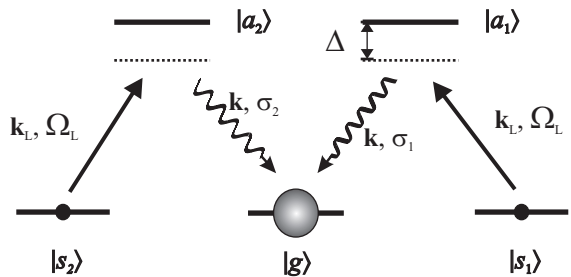}}
\caption{Double $\Lambda$-scheme for the generation of photons
  entangled in polarization.}
\label{fig.double.lambda}
\end{figure}

Before going into the description of particular experimental setups,
we discuss how to extend our formalism to generate photons
that are entangled in polarization. For this, we consider 
the double-$\Lambda$ scheme of Fig. \ref{fig.double.lambda},
in which there are two excited levels, $| s_1 \rangle$, $| s_2 \rangle$, and a ground
state level $| g \rangle$. The atom-photon mapping is described in
the same way, by defining the corresponding Holstein-Primakoff
operators $b_{1,j}$, $b_{2,j}$.
Assume that the $| s_1 \rangle$, $| s_2 \rangle$
states decay by emitting photons with different polarizations and 
creation operators $a^+_{1, \kk}$, $a^+_{2, \kk}$. If
the conditions for momentum conservation explained along this work are
fulfilled, then entangled spin waves will be mapped into entangled
photons in polarization. For example, consider the following initial
spin-wave state,
\begin{equation}
| \Psi \rangle_\aat
= \frac{1}{\sqrt{2}}
\left( b^+_{1, \KK_A} b^+_{2, \KK_B} + b^+_{1, \KK_B}
b^+_{2,\KK_A} \right) | 0 \rangle_\aat,
\label{entangled.spin.wave}
\end{equation}
where we have used the notation $b^+_{\sigma, \KK} =
(1/\sqrt{N})\sum_j e^{i \KK \rr_j} b^+_{\sigma,j}$.
After the atom-photon mapping, the following photonic state is created,
\begin{eqnarray}
& & \hspace{-0.75cm}
| \Phi \rangle_\pph = \frac{1}{\sqrt{2}} \nonumber \\
& & \hspace{-0.5cm}
\left( a^+_{1, \KK_A + \kk_\LL} a^+_{2,\KK_A + \kk_\LL} +
       a^+_{1, \KK_B + \kk_\LL} a^+_{2,\KK_A + \kk_\LL}
\right) | 0 \rangle_\pph .
\label{entangled.photons}
\end{eqnarray}
In this way, the ability to generate entangled spin-waves is
equivalent, by virtue of the atom-photon mapping, to the ability to
generate entangled photonic states.
We propose three experimental set-ups, where this is possible and discuss the
conditions that are required for the implementation of this idea.

\subsubsection{Trapped ions}

\begin{figure}[h]
\center
\resizebox{!}{!}{\includegraphics{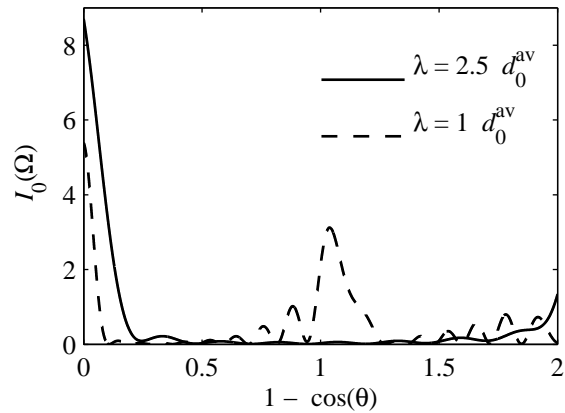}}
\caption{Angular photon distribution of the photonic mode emitted by
  the completely symmetric spin-wave in an ion chain with $N = 10$
  ions.}
\label{fig.ions.10}
\end{figure}

This system is specially appealing from the point of view of the
creation of entangled states, since quantum gates allow us to create
deterministically any collective state.
This idea has been demonstrated in the creation
of the $\KK = \00$ (W) state, in \cite{Häffner}. The first issue that
we have to deal with, is whether conditions for forward-scattering of
photons hold here. Usually ions are arranged in chains, with the peculiarity
that the distance between ions is not constant.
However, if condition $d^{\rm av}_0 < \lambda /2$,
with $d^{\rm av}_0$ the average distance, we still get light emission
in the forward--scattering cone only. We show this fact by
performing a calculation of the emission pattern of a single atomic
excitation in the $\KK = \00$ state, using the theory presented in the
previous sections, see Fig. \ref{fig.ions.10}.
Note that directionality is achieved even for a relatively small
number of ions ($N = 10$).
The main difficulty for the implementation of this idea with ions lies on
the fact that ion--ion distances are usually in the range of a few
$\mu m$, and thus condition $d^{\rm av}_0 < \lambda/2$ is not fulfilled
when considering optical wavelengths \cite{superradiance.ions}. 
A way out of this problem is
to use optical transitions which lie in the range of $\lambda
\gtrsim 5 \mu m$, like for example $\lambda$($^2$D$_{3/2}$ - $^2$P$_{1/2}$) = 10.8 $\mu$m in
Hg$^+$, or $\lambda$($^2$D$_{3/2}$ - $^2$D$_{5/2}$) = 12.5 $\mu$m in
Ba$^+$.
A particularly interesting initial atomic state is the spin-wave
(\ref{entangled.spin.wave}) with $\kk_A = \00$, and $\kk_B = -2
\kk_\LL$, (all vectors in the direction of the atomic chain), since
it leads to the emission of an entangled photon pair in the forward
and backward-emission directions.

\subsubsection{Ultracold atoms in optical lattices}

We have shown above that optical lattices are well suited to reach the
regime of directionallity in the emission of photons. The main issue
here, contrary to the case of trapped ions, is to find a way to create
efficiently the initial spin-wave states in a deterministic way.
Although one could think of peforming quantum gates between
ultracold neutral atoms to generate collective atomic states \cite{Jaksch,Mandel},
this procedure faces the difficulties of quantum computation in this system, like
for example, how to achieve single atom addressability.

\begin{figure}[h]
\center
\resizebox{!}{!}{\includegraphics{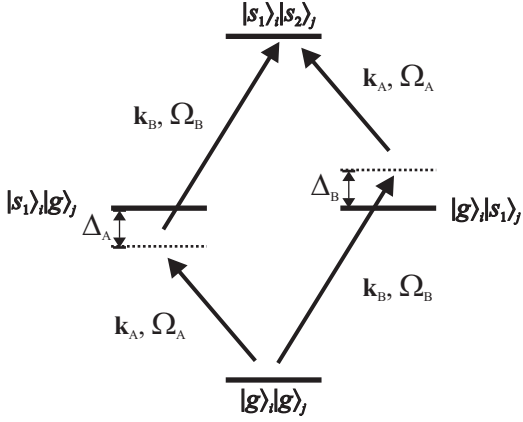}}
\caption{Scheme for the generation of entangled spin-waves by using
  the Rydberg blockade.}
\label{fig.rydberg}
\end{figure}

More efficiently, one could avoid the use of quantum gates by using
the dipole-blockade mechanism with Rydberg atoms, which allows us to
generate W-states, as well as states which emit Fock states with a
number $M$ of photons  \cite{LukinRydbergatoms}.
Interactions between excited atomic states, like those that take
place in Rydberg atoms, can be also used to generate photons entangled
in polarization.
This can be achieved in a single experimental step, without the need
for quantum gates, if the proper configuration of atomic interactions
is chosen.
As an example, consider the level configuration shown in
Fig. \ref{fig.rydberg}, and interactions between excited states such that
atoms in levels
$| s_1 \rangle$, $| s_2 \rangle$, interact strongly only if they are in the
same excited state, that is, $U_{11} = U_{22} = U$, but  $U_{12}=0$.
We apply two lasers with wavectors $\kk_{\AA,\BB}$ and Rabi
frequencies $\Omega_{\AA,\BB}$,
detuned with respect to the $| g \rangle$ -- $| s_{a,b} \rangle$ transition,
such that $\Delta_\AA = - \Delta_\BB = \Delta$. If condition $\Delta_{\AA,\BB} \gg
\Omega_{\AA,\BB}$ is fulfilled, then the lasers induce a two--photon
transition with Rabi frequency $\Omega_\textmd{eff} = \Omega_\AA
\Omega_\BB / \Delta$. Furthermore, if $\Omega_\textmd{eff} \ll U$,
states with two atoms in the same excited state are not populated.
Under these conditions there are two possible excitation channels,
depicted in Fig. \ref{fig.rydberg}, which give rise to the linear
combination (\ref{entangled.spin.wave}).

\subsubsection{Atomic ensembles at room temperature} 

The very same
techniques which can be applied to Rydberg atoms in an optical
lattice can also be used in the case of hot ensembles.
On the one hand, this setup has the advantage that atoms do not need to be
cooled and placed in an optical lattice. On the other hand, it can be
described by a statistical distribution of particles, and thus
suffers from the fact that high efficiency in the release of
photons is achieved under more severe conditions of particle
density and atom number, as discussed above. However, densities which
are high enough to fulfill the requirement ${\cal E} \ll 1$ have been
recently reported in \cite{Heidemann}.

\subsection{Photon up-conversion}
Finally, the ideas presented in this work find an intereseting
application in the efficient up-conversion of photons. 
For example, transitions between the states $| g \rangle$
and $| s \rangle$, are typically in the range of GHz. 
Application of
the $\Lambda$ scheme of Fig. \ref{fig.lambda}, provides a way for
up-conversion of the microwave photon to an optical
photon. 
Also, by using several
lasers, one could induce $N$-photon transitions from $|g\rangle$ to
$|a\rangle$, and emit a single high frequency photon. 
By using the ideas presented along this work, this could be
done in such a way that the up-converted photons 
are collimated and, thus, they can be efficiently collected.

\section{Acknowledgments}
We acknowledge interesting discussions with M. Lewenstein. Work
supported by the E.U. (SCALA), and the D.F.G. through the excellence
cluster Munich Advance Photonics (MAP).

\appendix

\section{Validity of neglecting propagation effects}
\label{appendix.propagation}
When performing the integration in $k$ in Eq. (\ref{define.matrix}), we
have to deal with the following factor inside the integration,
\begin{equation}
\frac{e^{i (\kk - \kk_\LL)(\rr_j - \rr_{j'})}}
{\left(-i(\omega_k - \omega_\LL) - \bar{J}_n^* \right)
 \left( i(\omega_k - \omega_\LL) - \bar{J}_{n'} \right)},
\label{factor}
\end{equation}
which can be evaluated in the following way,
\begin{eqnarray}
&&\frac{1}
{\left(-i(\omega_k - \omega_\LL) - \bar{J}_n^* \right)
 \left( i(\omega_k - \omega_\LL) - \bar{J}_{n'} \right)}
=
\nonumber \\
&&- \left(
\frac{1}{-i(\omega_k - \omega_\LL) - \bar{J}_n^*}
+
\frac{1}{i(\omega_k - \omega_\LL) - \bar{J}_{n'}}
\right)/(\bar{J}_n^* + \bar{J}_{n'})
\nonumber \\
&&
\approx -2 \pi \delta (\omega_k - \omega_L)
\frac{1}{\bar{J}^*_n + \bar{J}_{n'}} ,
\nonumber
\end{eqnarray}
where he have aproximated
\begin{eqnarray}
\RRe \left(
\frac{1}{-i(\omega_k - \omega_\LL - \Delta_n) - \Gamma_n/2}
\right)
\approx \pi \delta(\omega_k - \omega_\LL) .
\label{delta.approx}
\end{eqnarray}
This approximation is justified by the following argument.
The left-hand side of
(\ref{delta.approx}) is a function of $k$ with width $\Gamma_n/c$. The
latter has to be compared with the width of the exponential in
(\ref{factor}), which is, roughly $1/L$. Thus, the approximation holds
in the limit $\Gamma_n/c \ll 1/L$, that is, whenever the emission time
is shorter than the propagation of the photon through the sample,
something that is well justified in the trapping setups considered in
this work.
The energy shift can also be safely neglected, since condition
$\Delta_n \ll \omega_L$
is well justified by the atomic transitions used in cold atom setups.

\section{Multiphoton state emitted by
an atomic ensemble}
\label{appendix.multi.photon}
Our starting point is the atom-photon mapping defined in
Eq. (\ref{multi.photon.mapping}). By applying our method we get
\begin{eqnarray}
&& \hspace{-0.8cm} 
\rho = \frac{1}{M !} \frac{1}{N^M}
\langle
\left(
\sum_{j, \kk}
\frac{e^{i(\kk - \kk_\LL)\rr_j}}{i(\omega_k - \omega_\LL) - \Gamma_0/2} a^+_\kk
\right)^M
\nonumber \\
&& \hspace{-0.5cm}
| 0 \rangle_\pph \langle 0 |
\left(
\sum_{j', \kk'}
\frac{e^{-i(\kk' - \kk_\LL)\rr_{j'}}}{-i(\omega_{k'} - \omega_\LL) - \Gamma_0/2}a_{\kk'}
\right)^M
\rangle_{\rm mot} .
\label{lio}
\end{eqnarray}
After expanding the two parenthesis to the power of $M$ we can
reexpress this equation as a sum of $2 M$ products of exponentials of
the argument $i(\kk - \kk_\LL)\rr_j$. There are $\binom{N}{2 M}$ terms
such that all the random variables are different. These terms can be
summed up to a contribution which yields a pure state, such that 
$\rho$ is the sum of a pure and a mixed state,
\begin{equation}
\rho = \tilde{\rho}_{\rm pure} + \tilde{\rho},
\end{equation}
where $\tilde{\rho}_{\rm pure}$, $\tilde{\rho}$, 
are not normalized. The pure contribution has the form,
\begin{eqnarray}
&&\tilde{\rho}_{\rm pure} = \frac{1}{M !} \frac{1}{N^M}
\binom{N}{2 M}
\left(
\sum_{\kk}
\frac{g_k e^{-|\kk - \kk_\LL|^2 L^2/4}}{i(\omega_k - \omega_\LL) - \Gamma_0/2} a^+_\kk
\right)^M
\nonumber \\
&&| 0 \rangle_\pph \langle 0 |
\left(
\sum_{\kk'}
\frac{g_{k'} e^{-|\kk' - \kk_\LL|^2 L^2/4}}{-i(\omega_{k'} - \omega_\LL) - \Gamma_0/2}a_{\kk'}
\right)^M ,
\label{lio.2}
\end{eqnarray}
which is a pure state with $M$ photons.  
To normalize this state, we use the limit $\binom{N}{2 M} /N^M =
N^M, (N \gg M)$, and obtain the final result (\ref{multi.photon.dm}).
\section{3D Lattice, limit $\lambda \ll d_0$}
In the case of atoms in a 3D square lattice, in the limit $\lambda \ll
d_0$, there are in principle many diffraction peaks which contribute
to the emission pattern. In the following we show, that, at least for
the completely symmetric state, if the size of the system is large
enough, the forward-scattering contribution is the most important
one.

Recall the definition of the probability that the spin-wave $\00$
emits a photon in the $\mm$ Bragg scattering peak, $p^{[\mm]}_\00$ (\ref{def.pm}).
The forward-scattering contribution $\mm = \00$, is given by
\begin{equation}
p^{[\00]}_\00 = \frac{\bar{\Gamma}}{\Gamma_\00} \chi_{3D}.
\end{equation}
We consider for concreteness $\kk_\LL = k_\LL \hat{z}$.
To determine the probability of scattering out of the
forward-direction, 
${\cal E}$ 
we
calculate the sum over the contributions with $\mm \neq 0$,
\begin{eqnarray}
{\cal E} &=& \sum_{\mm \neq \00} p^{[\mm]}_\00 = \nonumber \\
& & \hspace{-1.5cm} \frac{\bar{\Gamma}}{\Gamma_\00}
\sum_{\mm \neq \00}
\frac{N}{4 \pi} \int d \Omega \ \pi^{3/2}
e^{- \left( \uu_\Omega - \hat{z} - \frac{\lambda}{d_0} \mm \right)^2
  (k_\LL d_0 N_x / 2)^2 }.
\end{eqnarray}
Considering the limit $k_\LL d_0 N_x/2 \gg 1$, the angular integral
yields,
\begin{equation}
{\cal E} = 
\sum_{\mm \neq \00}
\frac{\bar{\Gamma}}{\Gamma_\00} \pi^{3/2} \frac{N}{(k_\LL d_0 N_x)^2}
e^{- (1 - |\frac{\lambda}{d_0}\mm + \hat{z}|)^2 (k_\LL d_0 N_x / 2)^2
} .
\end{equation}
In the limit $d_0 \gg \lambda$, we can replace the sum by an
integration over $\mm$, and get the result,
\begin{equation}
{\cal E} = \frac{\bar{\Gamma}}{\Gamma_\00}.
\end{equation}
Together with the normalization condition $\sum_\mm p_\mm = 1$, this
leads to
\begin{equation}
{\cal E} = \frac{1}{1 + \chi_{3D}}, \ \ 
\frac{\Gamma_\00}{\bar{\Gamma}} = 1 + \chi_{3D} .
\end{equation}
Thus, as long as the optical thickness is large, the completely
symmetric state is superradiant, and photons are
collimated in the forward-scattering direction.
Note that the regime $\lambda
\ll d_0$ is very different than the directional regime, in the sense
that there is always a non-zero probability of emission out of the forward-scattering
cone. In the fully directional regime ($\lambda > 2 d_0$), on the contrary ${\cal E} = 0$,
up to finite size effects.
\section{Cycling transition}
\label{appendix.cycling}
\begin{figure}[h]
\center
\resizebox{!}{!}{\includegraphics{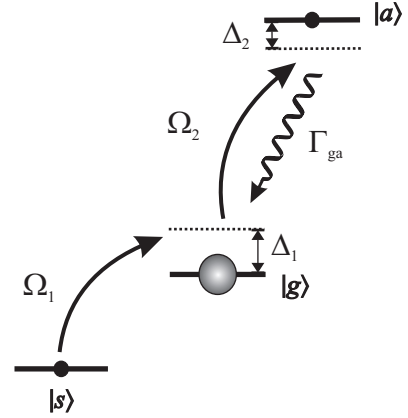}}
\caption{Level scheme for the implementation of the atom-photon
  mapping, avoiding incoherent processes.}
\label{fig.cycling.transition}
\end{figure}

Up to now we have neglected the decay form the auxiliary level $|
a \rangle$ back to the excited state $| s \rangle$. 
This process can be suppressed by considering, for example, a level configuration like the
one presented in Fig. \ref{fig.cycling.transition}.
$|s\rangle$ and $|g \rangle$ could be two hyperfine states of the
electronic ground state manifold, and $| a\rangle$ be chosen such that the
$|a \rangle$-$|g\rangle$ is a cycling transition. For example, these levels could be
found in $^{87}$Rb:  
$| s \rangle \to | {\rm S}_{1/2}, F = 1 , m_F = 1 \rangle$,  
$| g \rangle \to | {\rm S}_{1/2}, F = 2 , m_F = 2 \rangle$, and
$| a \rangle \to | {\rm P}_{1/2}, F = 3 , m_F = 3 \rangle$.
Spontaneous emission from $|a \rangle$ to $|s\rangle$ is forbidden by
selection rules on $m_F$, but atoms decay from $|a\rangle$ to
$|g\rangle$ with a rate $\Gamma_{\rm ga}$.
Fields with Rabi frequencies $\Omega_1$, $\Omega_2$, and detunings
$\Delta_1$, $\Delta_2$, are
coupled to the transitions $|s\rangle$-$|g\rangle$,
$|g\rangle$-$|s\rangle$, respectively.
Under some conditions, lasers induce a two-photon transition that is
equivalent to a $\Lambda$-configuration such as the one considered in
this work with $\Omega_\LL = \Omega_{\rm eff} = \Omega_1 \Omega_2 /
\Delta_1$. 

We consider two situations:

(i) $\Delta_2 \gg \Gamma_{\rm ga}$. 
The atom-photon mapping is possible under conditions:
$\Omega_{\rm eff} \ll \Delta_2$ (adiabatic elimination
of the level $| a \rangle$), $\sqrt{N}
\Omega_1  \ll \Delta_1$  (to avoid real transitions from $|s\rangle$
to $|g\rangle$), and 
$\Gamma_{\rm ga} (\Omega_{\rm eff}/\Delta_2)^2 \gg N \Gamma_{\rm ga}
(\Omega_2/\Delta)^2$, or equivalently, $\Omega_1 / \Delta_2 \gg
\sqrt{N}$, to avoid the dephasing induced by the transfer
of atoms from $| g \rangle$ to $| a \rangle$.

(ii) $\Delta_2 = 0$ (resonant case). Conditions:
$\Omega_{\rm eff} \ll \Gamma_{\rm ga}$ (adiabatic elimination of
$|a\rangle$). 
$\Omega_1 / \Delta_2 \gg \sqrt{N}$, and
$\Gamma_{\rm ga} (\Omega_{\rm eff} / \Gamma_{\rm ga})^2 \gg N
\Gamma_{\rm ga}(\Omega_2/\Delta_1)$, or equivalently, 
$\Omega_1^2 \gg N \Gamma_{\rm ga}^2$ (same reasons as (i)).

\newpage


\begin{thebibliography}{99}
\bibitem{note.rwa}
The rotating wave approximation is applied to derive
Eq. (\ref{H.lm}). However, counter-rotating terms must be included to
get correctly the dipole-dipole contribution to the master equation
coefficients. The inclusion of counter-rotating terms does not change
the mapping between Fock states of spin-waves to photons.
\bibitem{note.finite}
Since this result is obtained under the assumption of periodic
boundary conditions, subradiant states in finite lattices
will have a small non-zero decay rate due to finite size corrections.
\end{thebibliography}
\end{document}